\documentclass[fleqn,usenatbib,openleft]{mnras}

\usepackage{float}
\usepackage{graphicx}
\usepackage{hyperref}
\graphicspath{{./LEGACY/imgs/}}
\DeclareRobustCommand{\VAN}[3]{#2}
\let\VANthebibliography\thebibliography
\def\thebibliography{\DeclareRobustCommand{\VAN}[3]{##3}\VANthebibliography}

\usepackage{amsmath}	
\usepackage{subfig}
\def\Nlens{{\cal{N}}}
\def\mups{ \mu_{\scalebox{.9}{$\scriptscriptstyle PS$} }}
\def\mufs{ \mu_{\scalebox{.9}{$\scriptscriptstyle FS$} }}

\def\thetaE{\theta_{\rm E}}
\usepackage{xcolor}
\definecolor{mycolor}{rgb}{0,0,0}
\definecolor{mycolor2}{rgb}{0,0,0}
\definecolor{mycolor3}{rgb}{0,0,0}

\usepackage{enumitem}
\newcommand{\NewEntry}[2]{%
    #1 \par\noindent
    #2}
\title[Light Curve Calculations for Triple Microlensing Systems]{Light Curve Calculations for Triple Microlensing Systems}

\author[Kuang et al.]{
Renkun Kuang,$^{1,3}$\thanks{E-mail: krk18@mails.tsinghua.edu.cn}
Shude Mao,$^{1,2}$
Tianshu Wang,$^{4}$
Weicheng Zang,$^{1}$
Richard J. Long$^{1,5}$
\\
$^{1}$Department of Astronomy and Tsinghua Centre for Astrophysics, Tsinghua University, Beĳing 100084, China\\
$^{2}$National Astronomical Observatories, Chinese Academy of Sciences, 20A Datun Road, Chaoyang District, Beĳing 100012, China\\
$^{3}$Department of Engineering Physics, Tsinghua University, Beijing 100084, China\\
$^{4}$Department of Astrophysical Sciences, Princeton University, Princeton, NJ 08544, USA\\
$^{5}$Jodrell Bank Centre for Astrophysics, Department of Physics and Astronomy, The University of Manchester, Oxford Road, Manchester M13 9PL, UK}

\date{Accepted 2021 February 17. Received 2021 February 11; in original form 2020 August 27}

\pubyear{2021}

\begin{document}
\label{firstpage}
\pagerange{\pageref{firstpage}--\pageref{lastpage}}
\maketitle

\begin{abstract}
We present a method to compute the magnification of a finite source star lensed by a triple lens system based on the image boundary (contour integration) method. We describe a \textcolor{mycolor}{new} procedure to obtain \textcolor{mycolor}{continuous} image boundaries from solutions of the tenth-order polynomial obtained from the lens equation. \textcolor{mycolor}{Contour integration} is then applied to calculate the image areas within the image boundaries, which yields the magnification of a source \textcolor{mycolor}{with uniform brightness}. \textcolor{mycolor}{We extend the magnification calculation to limb-darkened stars approximated with a linear profile.} In principle, this method works for all multiple lens systems, not just triple lenses. We also include an adaptive sampling and interpolation method for calculating densely covered light curves. The C$++$ source code and a corresponding Python interface are publicly available.
\end{abstract}




\section{Introduction}

Gravitational microlensing has opened a unique window for probing extrasolar planets beyond the snowline (the minimum radius from a star at which water ice could have condensed) \citep{Shude1991,Andy1992, Gaudi2012, 2012RAA....12..947M, 2016ASSL..428..135G}. Among the 132 confirmed microlens planets, 12 were detected in a triple lens event\footnote{Information on the confirmed extrasolar planets is from \url{http://exoplanet.eu} as of Nov 20, 2020.}.

Triple lens systems are usually categorised into two groups, a host star plus two planets or a binary star system plus a single planet. Three two-planetary systems have been firmly established (OGLE-2006-BLG-109, \citealt{OB06109,OB06109_Dave}; OGLE-2012-BLG-0026, \citealt{OB120026,OB120026_AO,OB120026_Zhu}; OGLE-2018-BLG-1011, \citealt{OB181011}), and three likely candidates (OGLE-2014-BLG-1722, \citealt{OB141722}; OGLE-2018-BLG-0532, \citealt{OB180532}; KMT-2019-BLG-1953, \citealt{KB191953}). A planet in a binary star system (i.e., circum-binary planet) is another triple lens case. So far, six cases have been reported in the literature (OGLE-2006-BLG-284, \citealt{OB06284}; OGLE-2007-BLG-349, \citealt{OB07349}; OGLE-2008-BLG-092, \citealt{OB08092}; OGLE-2013-BLG-0341, \citealt{OB130341}; OGLE-2016-BLG-0613, \citealt{OB160613}; OGLE-2018-BLG-1700, \citealt{OB181700}).

The discovery rate of triple lens systems has nearly doubled since 2016, which is mainly attributed to the inauguration of the Korea Microlensing Telescope Network (KMTNet, \citealt{kim2016kmtnet}) in that year. With the continuous operation of KMTNet and the upcoming new facilities like EUCLID \citep{2010ASPC..430..266B, 2019ApJ...880L..32B} and WFIRST \citep{1808.02490}, we expect to encounter more triple lens events.

Analysing microlensing events is time consuming. Some binary events may even take years \citep{bozza2010microlensing}. The situation will be even worse when handling triple lens events because of the exponentially increasing parameter space. Thus, an efficient method is needed to model light curves for such systems. However, due to the inverse nature of solving the lens equation, and the finite source effect, the computation of magnification by a triple lens system is challenging. \textcolor{mycolor}{Three aspects need to be addressed, namely solving the lens equation numerically, dealing with complex caustic structures and image topologies, and handling finite source effects including limb-darkening. Our approach is explained in the next section.}

\textcolor{mycolor}{Currently, there are two main schemes for modelling triple lens events. The first approach is based on the ``binary superposition'' approximation method \citep{2001MNRAS.328..986H, 2002MNRAS.335..159R, 2005ApJ...629.1102H, OB120026}, while the second is based on the ray-shooting method \citep{1986A&A...166...36K, 1987SchneiderRayshooting}. Neither scheme is completely satisfactory however.}

For some triple events, their light curves can be approximated as a superposition of two binary light curves. Nevertheless, the superposition method is not always valid and sometimes the detectability of a second planet will be suppressed by the presence of the first planet \citep{2014ApJ...794...53Z, 2014MNRAS.437.4006S}. \textcolor{mycolor}{\textcolor{mycolor2}{In} ``binary superposition'', \textcolor{mycolor2}{ several methods may be used to calculate the binary light curves, including }the contour integration method \citep{gould1997stokes, dominik1998robust}. \textcolor{mycolor2}{In this work, the general contour integration method introduced in \cite{gould1997stokes} is implemented in the triple-lens scenario.}}

\textcolor{mycolor2}{To calculate magnification for finite sources, \cite{2000Vermaak} started from finding the image positions corresponding to the source centre, then used a recursive flood-fill algorithm to check whether neighbouring integration elements in the image plane can be mapped onto the source through the lens equation.} \cite{dong2006planetary, dong2009microlensing} and \cite{2014ApJ...782...47P} advocated the mapmaking method, which is a hybrid of the ray-shooting and contour integration methods. \cite{bennett2010efficient} proposed a method for modelling high magnification events for multiple-lens microlensing events, based on the image centred ray-shooting approach of \cite{bennett1996detecting}.

In contrast, \cite{mediavilla2006fast, mediavilla2011new} proposed an approach based on inverse polygon mapping to compute the magnification maps. 
\textcolor{mycolor2}{For ray-shooting methods,} if the source radius is small or close to the caustics, high density light rays are needed. \textcolor{mycolor}{Other than for validation purposes, in this work we choose not to use ray-shooting for triple microlensing light curve calculations.}


The purpose of this paper is to present a method for calculating the magnification of a limb-darkened finite source lensed by a triple lens system. \textcolor{mycolor}{Our approach, overviewed in \S2, is based on calculating image areas by contour integration but with an alternative procedure for obtaining the image boundaries.} In \S3, we introduce the complex lens equation \textcolor{mycolor}{and our notations}, while in \S4, we present the details of our method. In \S5 we present the results. Finally, a short summary is given in \S6.

\section{Approach}
\textcolor{mycolor}{Modelling lensed light curves is achieved by setting up a representation of the lensing system and then repeatedly moving a source across the lensing system and using its magnifying effect to generate model light curves. Every model light curve is compared with the observed light curve and various lensing parameters are adjusted until a best-fitting curve is found. The focus of this paper is on the magnification calculation required for a limb-darkened, finite-sized source.}

Real source stars have finite sizes, and this causes their light curves to be significantly modified in high magnification regions \citep{witt1994can, 1994ApJ...421L..71G}. Finite source effects are essential for binary and triple lens systems because they are related to the mass ratios of the lens components, and can lead to measurements of the angular Einstein radius $\theta_{\rm E}$ when combined with knowledge of the angular source radius $\theta_*$ \citep[e.g.,][]{Yoo2004}. Finite source effects are particularly important when the source crosses a caustic (where the magnification diverges to infinity) or comes close to a cusp caustic \citep{witt1994can,1994ApJ...421L..71G,Nemiroff1994}. In these cases, the source cannot be regarded as point-like, and the observed magnification is an average of the magnification pattern over the face of the source. \textcolor{mycolor}{The surface brightness of a star is however not uniform \citep{1921Milne}. This is known as the limb-darkening effect, which also affects the microlensing magnification \citep{1996ApJ...464..212G, 1999Spectrophotometric, 2003measurelimb, 2005MNRAS.361..300D}. Limb-darkening effect needs to be considered to model precisely observed light curves, e.g., as in the first limb-darkening measurement by microlensing, the event MACHO 1997-BLG-28 \citep{2001firstlimbdetect}. It is thus crucial to have a reliable and efficient way to calculate the magnification of a limb-darkened, finite size source for triple lens systems.}

\textcolor{mycolor}{
For every position of the finite source, the lens equation has to be solved at multiple points around the source boundary so that the boundaries of the images can be determined. Gravitational lensing preserves surface brightness \citep{1973gravbook}. For a source with uniform surface brightness, the magnification due to lensing is equal to the ratio of the total image area and the source area. Usually, a two dimensional image area is computed with double integrals but these can  be converted into a line integral according to Stokes' theorem. This is the contour integration technique \citep{1987contour, dominik1993, dominik1995, dominik1998robust, gould1997stokes, bozza2010microlensing,bozza2018vbbinarylensing} that we use in our work.
}

\textcolor{mycolor}{The triple lens equation has to be solved numerically.} The lens equation for a single lens can be easily solved: the two image positions and magnifications can be derived analytically \citep{Einstein506,paczynski1986gravitational}. The lens equation for a binary lens is considerably more complicated, as it is no longer analytically tractable. The binary lens equation can be transformed into a fifth-order complex polynomial \citep{witt1990investigation}. \citet{skowron2012general} provided an algorithm to solve complex polynomial equations, and we use this algorithm in our modelling. It has been used elsewhere in a public package named VBBinaryLensing to calculate microlensing light curves for binary lens systems \citep{bozza2010microlensing, bozza2018vbbinarylensing}. They found that $\sim 80\%$ of computer time is spent in the root finding routine \citep{bozza2010microlensing}. \textcolor{mycolor}{This usage would increase for the tenth order polynomial for a triple lens system if they were to extend their method to the triple lens scenario.}

\textcolor{mycolor}{
We choose not to use the image boundaries obtaining method in VBBinaryLensing as we believe it would require significant effort to extend it to the triple lens scenario. We have implemented an alternative strategy for determining image boundaries which we describe in \S\ref{sec::conn}, and then compare with previous methods in \S\ref{sec::comp}.
}

\section{General Concepts}
\textcolor{mycolor}{In this section we introduce the complex lens equation and the parametrizations we use.}
\subsection{Lens equation for N point lenses}
Using complex notations, the $\Nlens$ point lens equation can be written as \citep{witt1990investigation}
\begin{equation}
\label{lensequ}
    \zeta = z + f(\overline{z}), \;\; f(\overline{z})\equiv - \sum\limits_{j=1}^{\Nlens}\frac{m_j}{\overline{z}-\overline{z_j}},
\end{equation}
where $\zeta=y_1+i\,y_2$ is the source position, $z=x_1+i\,x_2$ is the corresponding image position. $m_j,\,z_j$ are the fractional mass and position of $j$-th lens, $\sum_j m_j = 1$, $\overline{z}$ and $\overline{z_j}$ are the complex conjugates of $z$ and $z_j$. \textcolor{mycolor}{If a point $z$ in the lens plane satisfies the lens equation, it will map back to the source position $\zeta$ through the lens equation, in this case we call $z$ a true image of $\zeta$.} \textcolor{mycolor2}{The position coordinates $\zeta,\, z,\, z_j$ are in units of the angular Einstein radius $\thetaE$ of the lensing system,}
\begin{equation}
\thetaE = \sqrt{\frac{4GM}{c^2}\frac{D_{ls}}{D_lD_s}}\;,
\end{equation}
\textcolor{mycolor}{
where $D_l,\;D_s$ are the angular diameter distances of the observer to the lens and to the source, $D_{ls}$ is the angular diameter distance between the lens plane and the source plane, $M$ is the total mass of the lens system, $c$ is the speed of light, $G$ is the gravitational constant.}

Taking the conjugate of equation (\ref{lensequ}), we obtain an expression for $\overline{z}$, which one can substitute back into the original lens equation to obtain an $\Nlens^2+1$ order complex polynomial in $z$ only (\citealt{witt1990investigation}). So the maximum number of roots (images) cannot exceed $\Nlens^2+1$. In fact, some of these roots are not \textcolor{mycolor2}{true images}, \textcolor{mycolor}{i.e., although they satisfy the complex polynomial, they do not satisfy the original lens equation, in this case, they are called false images. In \cite{bozza2010microlensing}, such images are called ghost images.} It has been shown that the true upper limit of true images is $5(\Nlens - 1)$ \citep{2001astro.ph..3463R, 2003astro.ph..5166R, 2004math......1188K}. Notice that for $\Nlens>3$, there must be false images from solving the polynomial for any given source position. \textcolor{mycolor}{We will introduce how to distinguish true and false images in \S\ref{sec::cri_truesolution}.}

For each image position $z$, its magnification ($\mu$) is related to the Jacobian $J$ by
\begin{equation}
    \mu=J^{-1},\;\; J = 1 - f^{\prime}(\overline{z}) \overline{f^\prime(\overline{z})}, \;\; f^\prime(\overline{z}) = \text{d}f(\overline{z})/d\overline{z},
\end{equation}
where $f(\overline{z})$ is defined in equation (\ref{lensequ}). \textcolor{mycolor}{Some image positions will lead to $J = 0$ with the magnification $\mu$ becoming infinite. These image positions constitute ``critical curves'' in the lens plane, with the corresponding source positions forming ``caustics'' in the source plane.} The total magnification $\mups$ can be obtained by summing the inverse Jacobian determinant for each true image $z_I$ $(I=1,\cdot\cdot\cdot, N_{\rm im})$, corresponding to the source $\zeta$
\begin{equation}
    \mups = \sum\limits_{I=1}^{N_{\rm im}}\frac{1}{|J(z_I)|},
\end{equation}
where $N_{\rm im}$ is the total number of true images.

\subsection{Triple lens systems: parametrization}
\begin{figure}
\includegraphics[width=\linewidth,keepaspectratio]{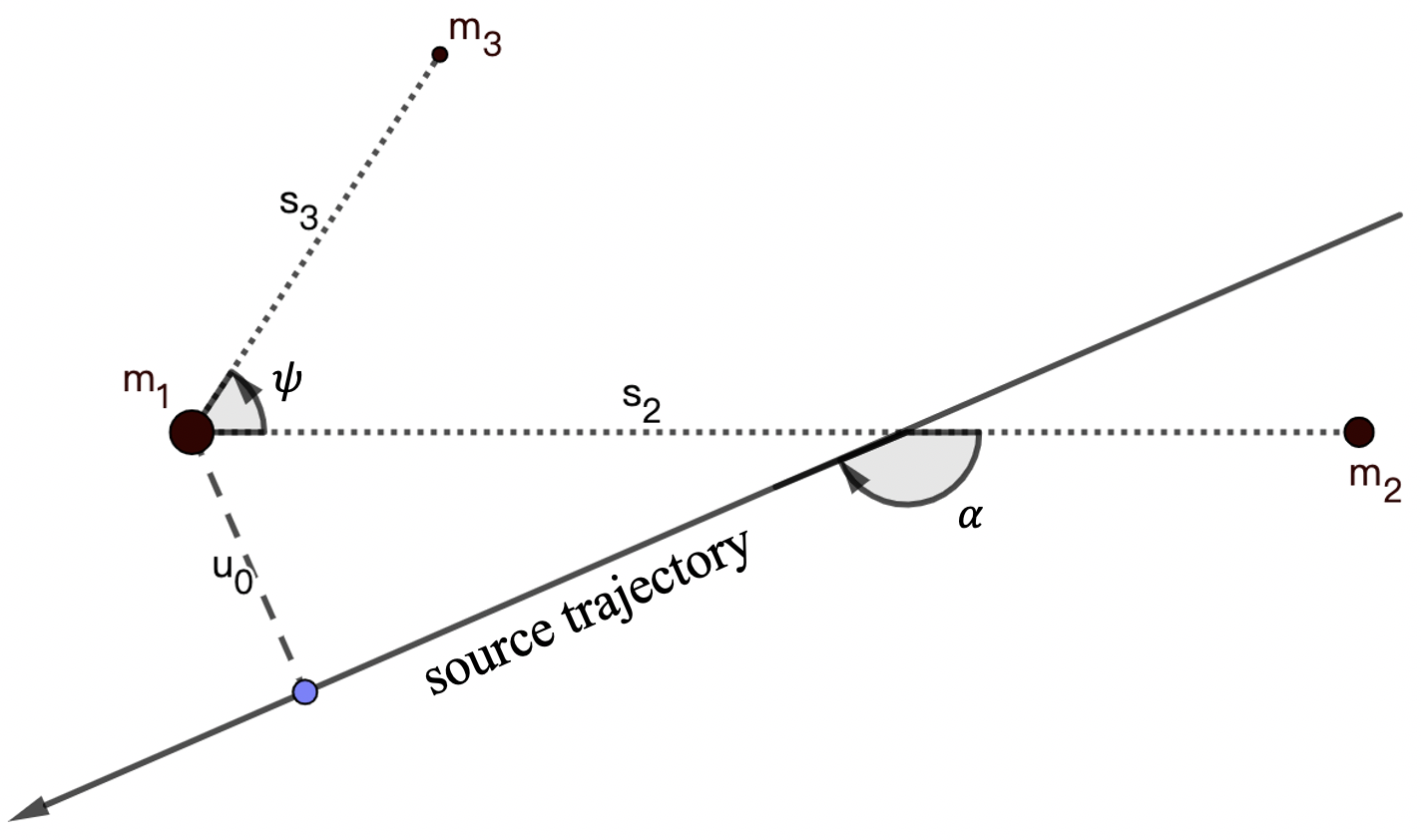}
\caption{Schematic diagram of a triple lens system. \textcolor{mycolor}{$m_1,\,m_2$ and $m_3$ are the fractional masses of the three lenses ($m_1+m_2+m_3=1$, \textcolor{mycolor2}{$m_1\geq m_2 \geq m_3$}). The separations of $m_2$, $m_3$ relative to $m_1$ are labelled as $s_2$ and $s_3$, $m_1$ and $m_2$ are on the horizontal axis, $\psi$ is the direction angle of $m_3$ relative to the horizontal axis. The source trajectory is parametrized by the trajectory angle $\alpha$ and the impact parameter $u_0$.}}
\label{fig:graph}
\end{figure}

We ignore the source baseline flux \textcolor{mycolor}{(source flux before microlensing event happens, \citealt{2000MNRASBaseline})} and blending flux \textcolor{mycolor}{(flux that is not physically associated with the lensed source, \citealt{1995ApJLblending, 2007MNRAS.380..805S})}, which are easy to model linearly. With this simplification, there are ten parameters in modelling triple lens light curve, five lens parameters ($s_2$, $q_2$, $s_3$, $q_3$, $\psi$) and four trajectory parameters ($t_0$, $u_0$, $t_{\rm E}$, $\alpha$), plus the source radius $\rho$. $\rho =\theta_{*}/\thetaE$, where $\theta_{*}$ and $\thetaE$ are the angular source radius and the angular Einstein ring radius of the lensing system, respectively. $s_2$ and $q_2$ are the separation and mass ratio between the first and second lenses, i.e., $q_2 = m_2/m_1$. Similarly, $s_3$ and $q_3$ are the separation and mass ratio between the first and third lenses. $\psi$ is the orientation angle of the third body. $t_0$ is the time when the source is closest to the primary mass $m_1$. The impact parameter $u_0$ is the primary lens-source separation at $t_0$. $s_2,\,s_3$ and $u_0$ are in units of $\thetaE$. The Einstein timescale $t_{\rm E}$ controls the duration of the event. $\alpha$ is the source trajectory angle with respect to the $m_1$-$m_2$ axis. \textcolor{mycolor2}{We note that the number of parameters in many triple events is even higher due to non-trivial lens motion (parallax or orbital motion effects). In addition, for a finite source, limb-darkening parameters (filter-specific) are relevant too. To start, we focus on the magnification of a uniform brightness star.} A graphical illustration of the triple lens system (similar to Fig. 1 of \citealt{OB160613}) is shown as Fig. \ref{fig:graph}.

\textcolor{mycolor2}{Since most of the observed triple microlensing events have a third, low-mass body (either a second planet or a planet in a binary)}, we choose the origin of the coordinate system to be the centre of mass of the first two masses. Thus the conversion from ($s_2$, $q_2$, $s_3$, $q_3$, $\psi$) to $m_j, z_j$ is as follows
\begin{equation}
\begin{aligned}
  m_1 &= 1 / (1 + q_2 + q_3),\\
  m_2 &= q_2\; m_1,\\
  m_3 &= q_3\; m_1 = 1 - m_1 - m_2,\\
  z_1 &= -q_2\;s_2/ (1 + q_2) + i\,0,\\
  z_2 &= s_2/(1 + q_2) + i\,0,\\
  z_3 &= -q_2\;s_2/ (1 + q_2) + s_3\cos(\psi) + i\,s_3\sin(\psi).
\end{aligned}
\end{equation}

\textcolor{mycolor}{We note that whichever coordinate system is chosen, magnification can still be calculated using this method. The input parameters are $m_j,\,z_j$, $\rho$, and $\zeta$, and the output is the magnification at $\zeta$.}


\subsection{Finite sources: parametrization}
\label{sec::srcbound}
For a circular source with radius $\rho$, centred at $\zeta^{(c)} = y_1^{(c)} + i\,y_2^{(c)}$, its boundary can be represented as
\begin{equation}
\zeta(\theta) = \zeta^{(c)} + \rho e^{i\theta},\;\; \theta \in [0,2\pi].
\end{equation}

In practice, \textcolor{mycolor}{we approximate the circular source boundary by a polygon with $n$ different vertices $\zeta(\theta_k)$}, where $\theta_0 < \theta_1 < \cdot\cdot\cdot < \theta_k < \cdot\cdot\cdot < \theta_n = \theta_0 + 2\pi$. The images of the source are distorted by the lens system (see Fig. \ref{fig:topo}). \textcolor{mycolor}{The $\theta_{k}$ are not necessarily sampled at equal intervals}. For each $\theta_k$, we need to solve the corresponding lens equation to obtain \textcolor{mycolor}{both true and \textcolor{mycolor2}{false images}, and attach them to image tracks. Then we pick the true image segments out to obtain the true image boundaries.} Finally, we can apply Stokes' theorem to obtain the enclosed area of these image boundaries.

\section{Finite source magnification for triple lenses}
\begin{figure*}
\includegraphics[width=1\linewidth]{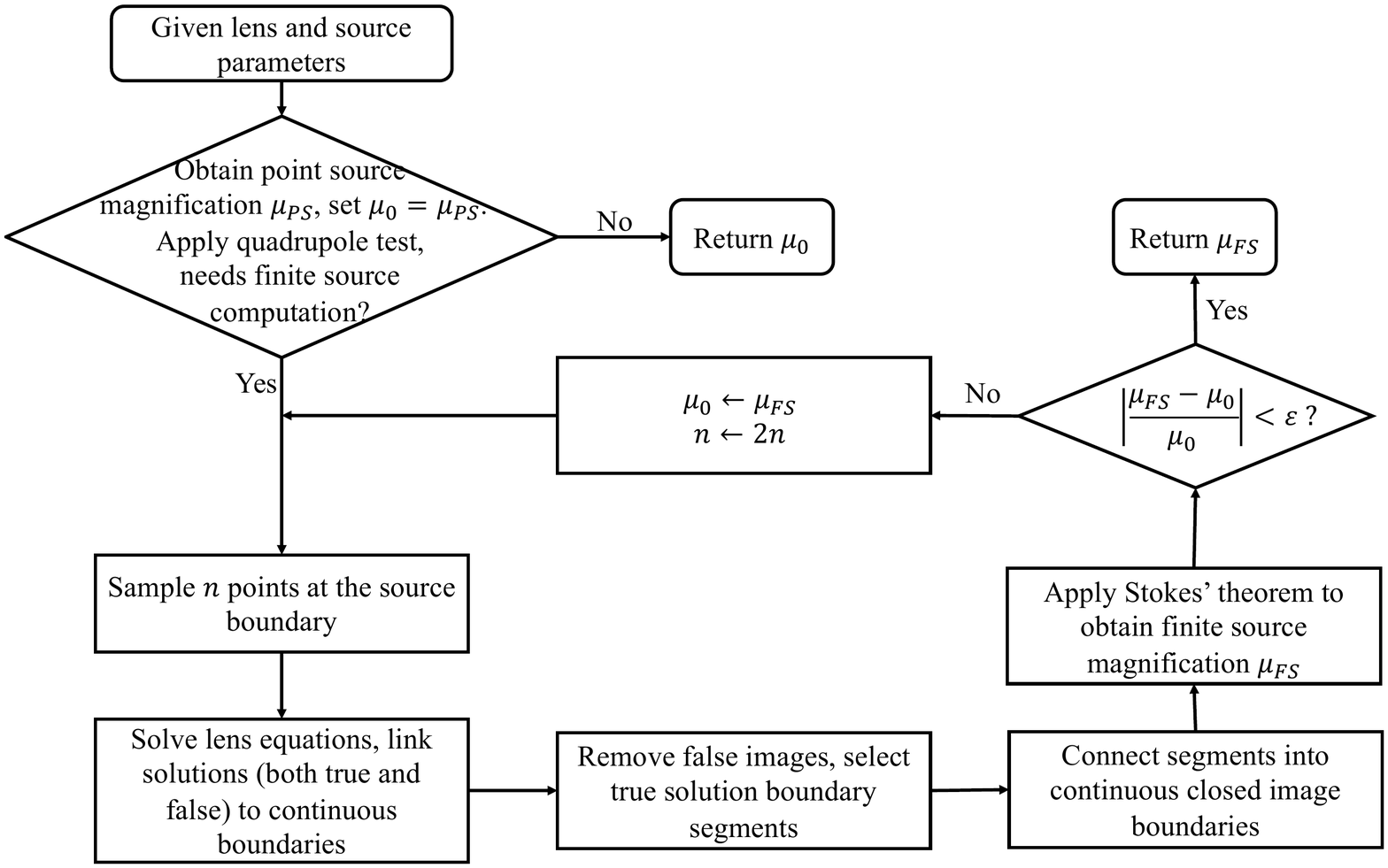}
\vspace{-1.2cm}
\caption{A flowchart describing our procedure \textcolor{mycolor}{to calculate the magnification of a uniform brightness star}. \textcolor{mycolor}{The three procedures along the bottom of the flowchart (solve the lens equations, remove false images, and connect segments) are described in detail in \S \ref{sec::conn}}.
}
\label{fig:flowchart}
\end{figure*}

\textcolor{mycolor}{Before applying Stokes' theorem to calculate the magnification of a uniform brightness star, one needs to obtain continuous image boundaries.} In \S\ref{sec::conn}, we discuss the new method we have developed to connect the image boundaries. In \S\ref{sec::quad} we have also adopted Bozza's quadrupole test \citep{bozza2018vbbinarylensing} to decide whether point source magnification $\mups$ is sufficient to approximate the magnification $\mufs$ of a uniform brightness star. The flowchart in Fig. \ref{fig:flowchart} shows the major stages we execute. \textcolor{mycolor}{Once we can calculate the magnification of a uniform brightness star, we can model limb-darkening light curves by regarding the source star as a set of annuli weighted by the radial limb-darkening profile, as introduced in \S\ref{sec::limb}.}



\subsection{Topology of image boundaries}
\label{sec:topo}
\begin{figure*}
\begin{center}
\includegraphics[width=0.48\linewidth,keepaspectratio]{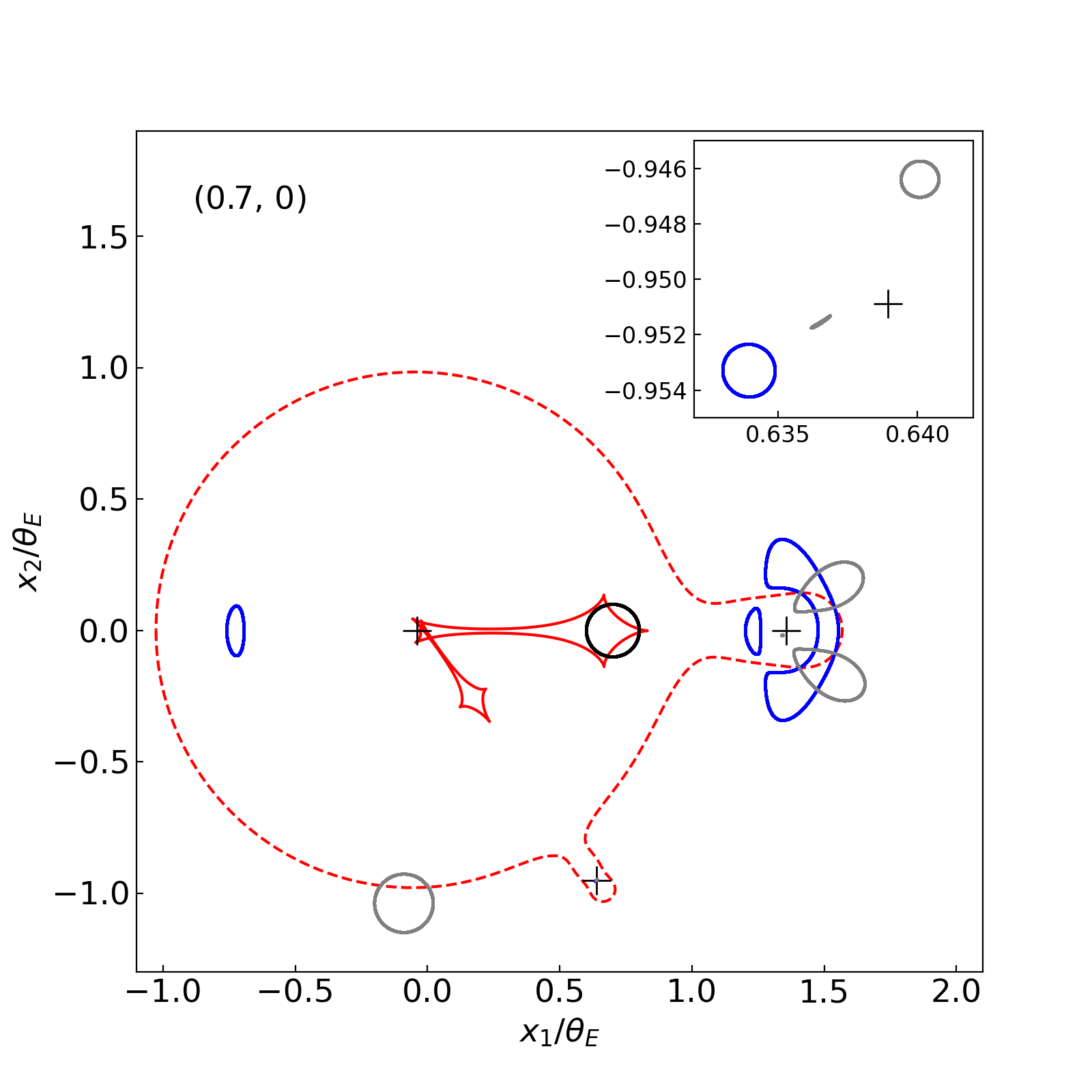}
\includegraphics[width=0.48\linewidth, keepaspectratio]{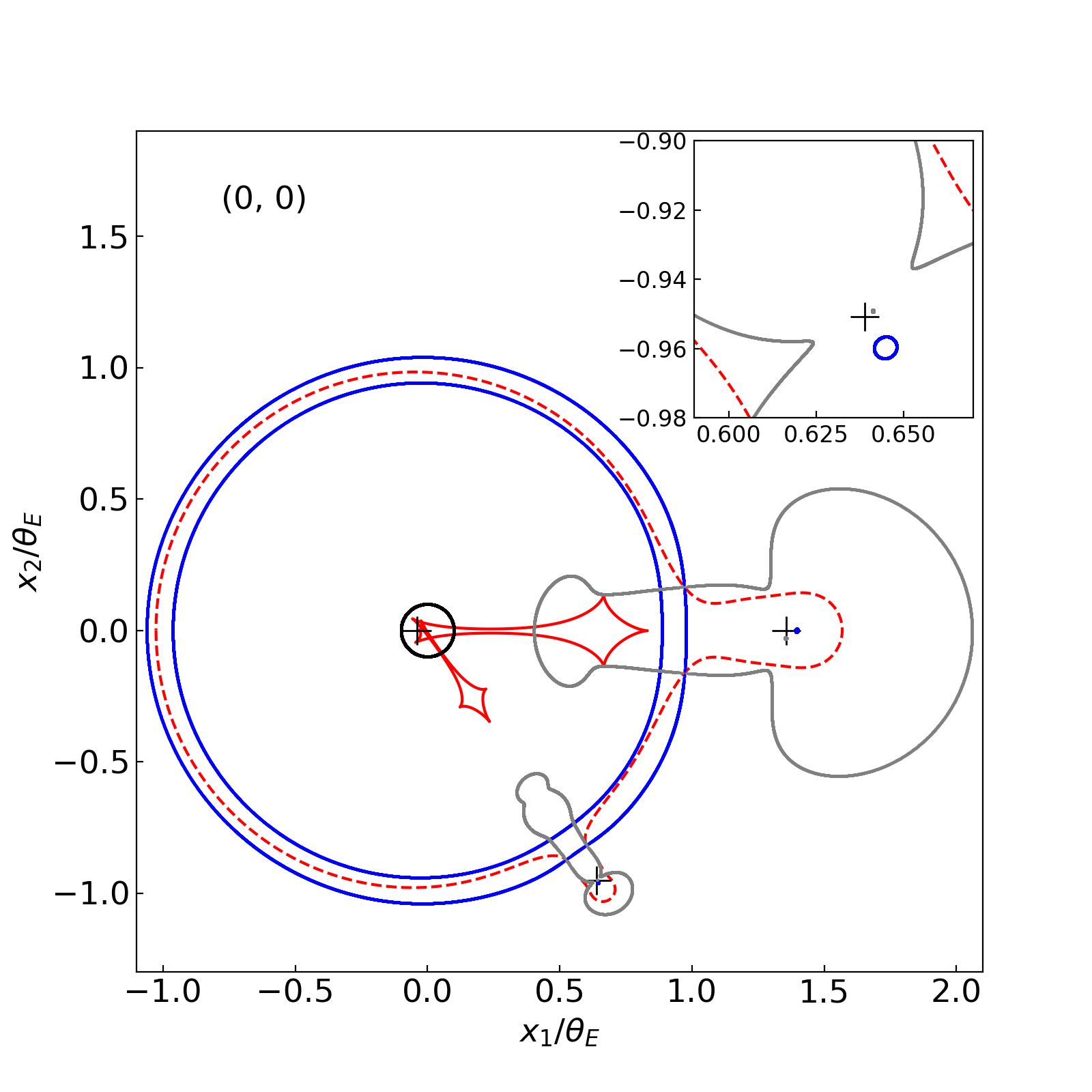} 
\end{center}
\vspace{-0.7cm}
\caption{Illustrations of image boundary topologies formed by the roots of the complex polynomial for a triple lens system. Parameters of this system are introduced in \S \ref{sec:topo}. The three plus signs indicate the lens positions, the red solid and dashed curves show the caustic and the critical curve. The circular source is indicated by the black circle ($\rho = 0.1$). \textcolor{mycolor}{The blue curves are true image boundaries while the grey curves are false image boundaries arising from the tenth-order polynomial. \textcolor{mycolor2}{Insets show the detail around the third mass.} In the left panel, the source is at (0.7, 0) while in the right panel it is at (0, 0) in the source plane. In the right panel, the two primary image boundaries are nested, forming a ``ring-like'' image, which needs special care when calculating the enclosed image area inside that ``ring''.}
}
\label{fig:topo}
\end{figure*}

\textcolor{mycolor}{We first illustrate} the topology of both true and false image boundaries for the circular source lensed by a triple system in Fig. \ref{fig:topo}. We adopt the parameters of the triple lens solution ``Sol C (wide)'' \textcolor{mycolor}{of microlensing event  OGLE-2016-BLG-0613 \citep{OB160613}}. According to their Table 3, $t_0 = 7494.153$;  $u_0 = 0.021$; $t_{\rm E} = 74.62$; $s_2 = 1.396$; $q_2 = 0.029$;  $\alpha = 2.948$; $s_3 = 1.168$; $q_3 = 3.27\times 10^{-3}$; $\psi = 5.332$. Instead of $\rho = 2.2\times 10^{-4}$ in their solution, to visualise better the image boundaries, we set $\rho = 0.1$. \textcolor{mycolor}{Two} source centres are shown. \textcolor{mycolor}{In the right panel, the two primary image boundaries are nested, forming a ``ring-like'' image, which needs special care when calculating the enclosed image areas.}

\textcolor{mycolor}{
In Fig. \ref{fig:topo}, the blue curves are true image boundaries, and the grey curves are false image boundaries arising from the tenth-order polynomial. \textcolor{mycolor2}{In the left panel, there are four true image boundaries (three large ones plus a small one near the third lens mass, in blue) and six false image boundaries (three large ones and a small one near the second mass plus two small ones near the third mass, in gray). In the right panel, there are four true image boundaries (two large ones form the ring structure, and the smaller two are close to the second and third lens masses) and four false image boundaries (two large mushroom-like images and two small circular boundaries close to the second and third lens masses).} Such topological figures merely give a preliminary impression about the shape and configuration of the image boundaries. The plotted data are created by solving the lens equation to give points along the boundary. Quantitative information, i.e., enclosed areas can not be calculated using line integrals until these points are connected in order (clockwise or anti-clockwise).
}

\subsection{Connecting image points to obtain continuous image boundaries}
\label{sec::conn}
\textcolor{mycolor}{
We use the configuration in the left panel of Fig. \ref{fig:topo} as an example, i.e., the source centred at $(0.7, 0)$, to illustrate how we construct the continuous image boundaries.}

The outer limb of the source is approximated by a polygon as described in \S\ref{sec::srcbound}. \textcolor{mycolor}{
We first initialise $\Nlens^2+1$ linked lists to store images from solving lens equations, here $\Nlens$ is the number of lenses. At each source position $\zeta({\theta_k})$, $k=0,1, \cdot\cdot\cdot, n$, we solve the corresponding polynomial lens equation, which will generate $\Nlens^2+1$ solutions. Each solution is attached to a linked list, depending on its distance from the tails of linked lists.}

\begin{figure}
\begin{center}
\includegraphics[width=\linewidth,keepaspectratio]{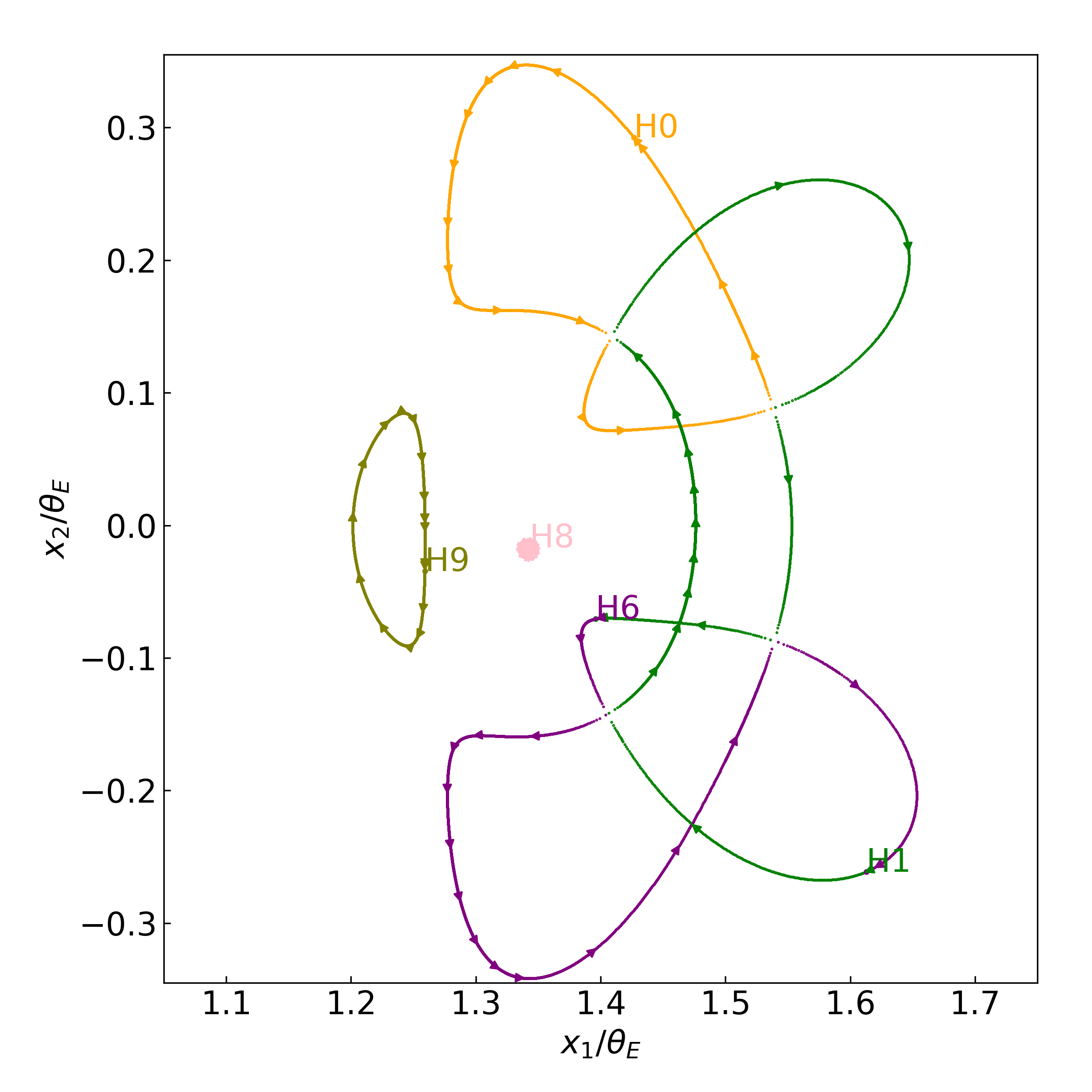}
\end{center}
\vspace{-0.5cm}
\caption{Illustrations of the $\Nlens^2+1 = 10$ linked lists after triple lens equation solving and preliminary points-connecting. Parameters are the same as the left panel of Fig. \ref{fig:topo}. Points in different linked lists are plotted with different colours, and the head point of the $i$-th linked list is labelled as $H\{i\}$. Arrows are added to indicate the direction in which points are linked inside each linked list, from head to tail. We just show part of all ten linked lists to help visualisation.
}
\label{fig:n+1_linked_list}
\end{figure}

\textcolor{mycolor}{
After the above process, we will obtain $\Nlens^2+1$ linked lists, which store the image points. Each linked list contains the same number of points. As Fig. \ref{fig:n+1_linked_list} shows, points in different linked lists are plotted with different colours, and the head point of the $i$-th linked list is labelled as $H\{i\}$. We add arrows to indicate the direction in which points are linked from the head to the tail of a linked list. For triple lens, i.e., $\Nlens = 3$, there are ten linked lists. In Fig. \ref{fig:n+1_linked_list}, we show only a subset of those ten linked lists to help visualisation.
}

\textcolor{mycolor}{
Connecting points at this stage is just a preliminary procedure, since it is not guaranteed that every point will be attached to the right place. Some linked lists contain only true images (e.g., the $9$-th list in Fig. \ref{fig:n+1_linked_list}), some contain only false images (e.g., the $8$-th list in Fig. \ref{fig:n+1_linked_list}), while others may contain both (e.g., the $0,\,1,\,6$-th list in Fig. \ref{fig:n+1_linked_list}). Mixing will happen, especially during caustics crossing, when two image boundaries are very close to, or intersect with each other. In such cases, we need to go through several steps to obtain true image boundaries which contribute to the total magnification. We want to emphasise that, \textcolor{mycolor2}{false images} actually help us to connect true image points correctly later in the process. False images can be considered as ``bridges'' which link different true image segments together. If we do not use the position information of the \textcolor{mycolor2}{false images}, it would be hard to link all the true image points into continuous boundaries. Because the number of \textcolor{mycolor2}{true images} would change over the source boundary \textcolor{mycolor2}{(as the sampled point crosses the caustics)}.}



\begin{figure}
\begin{center}
\includegraphics[width=\linewidth,keepaspectratio]{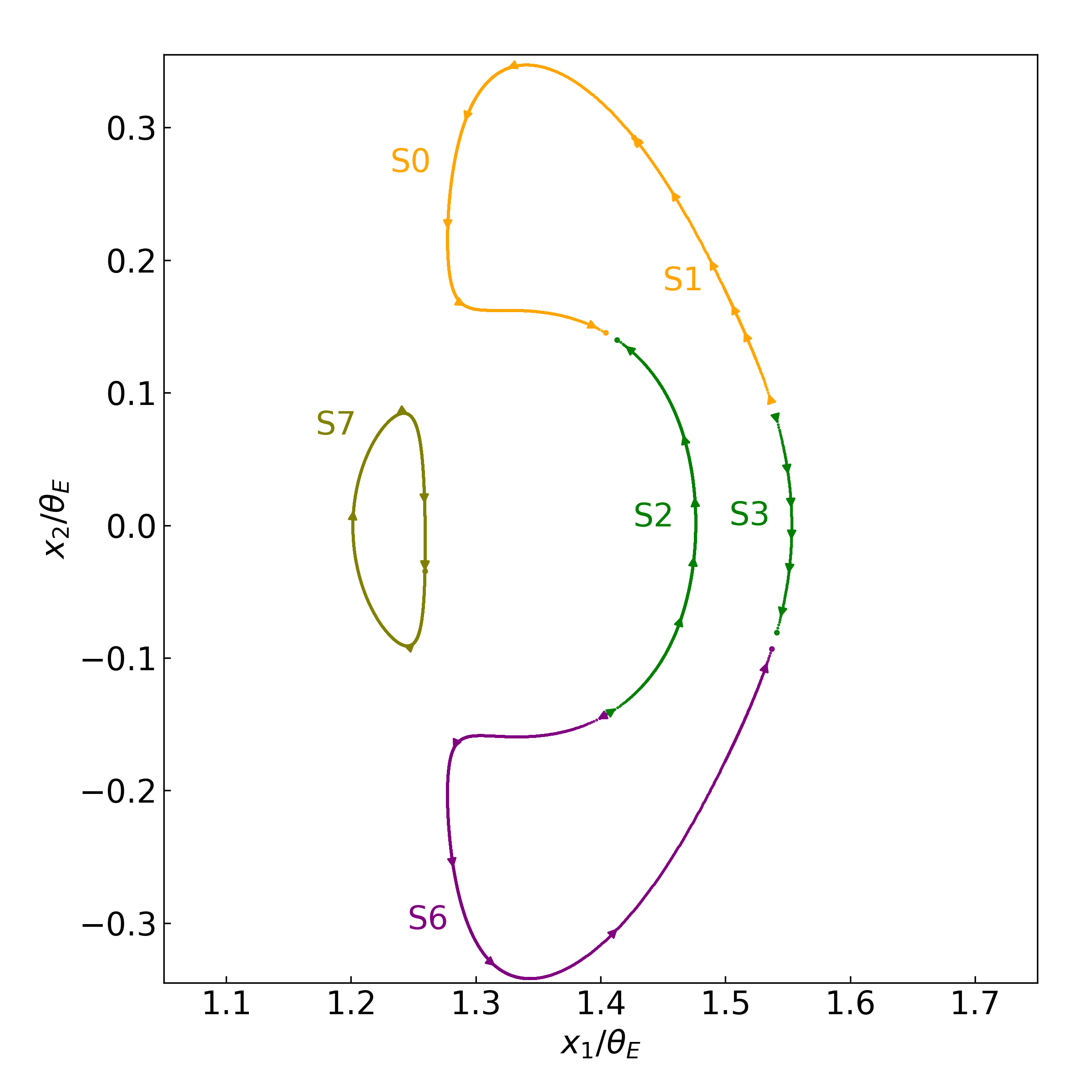}
\end{center}
\vspace{-0.5cm}
  \caption{
Demonstration of true image solution segments after removing false image points in Fig. \ref{fig:n+1_linked_list}. Different segments are labelled with $S\{i\}$. \textcolor{mycolor3}{The colour} of each segment \textcolor{mycolor3}{is inherited} from their original linked list in Fig. \ref{fig:n+1_linked_list}. Arrows \textcolor{mycolor3}{indicate} the direction in which points are linked from the head to the tail of each individual segment.
}
\label{fig:segms}
\end{figure}

We use the lens equation to check whether a root is a true image \textcolor{mycolor}{(as described in \S \ref{sec::cri_truesolution})}. We then remove all \textcolor{mycolor2}{false images} from the linked lists. \textcolor{mycolor}{After this, one usually breaks the $\Nlens^2+1$ linked lists obtained previously into several image segments, as shown in Fig. \ref{fig:segms}. Notice that the number of resulting segments is not necessarily $\Nlens^2+1$ anymore. Different segments are labelled as $S\{i\}$. Arrows indicate the direction in which points are linked in each segment from head to tail. The colour of each segment is inherited from the original linked list in Fig. \ref{fig:n+1_linked_list}. The last step before applying Stokes' theorem is to connect those image segments into closed continuous image boundaries}.

To do this, we first check whether any given segment is closed by evaluating the distance $d$ between the head and tail of a segment (we choose a threshold $10^{-5}$, \textcolor{mycolor}{which will be explained in more detail in \S \ref{sec::cri_vicinity}}). If $d\leq10^{-5}$, then this completes our task for the current segment (\textcolor{mycolor}{see segment ``S7'' in Fig. \ref{fig:segms}}). We move to the next segment until all segments are processed. If the current segment is not closed, we first check whether its tail connects with the head or tail of other image segments by checking whether they have identical image positions. If not, we also conduct the check for the head of the current segment (\textcolor{mycolor}{see the segment ``S0'' and ``S1'' in Fig. \ref{fig:segms}, the head of ``S0'' is connected with the tail of ``S1''}). Another possibility to connect two segments occurs when the \textcolor{mycolor}{image} boundary crosses the critical curve (\textcolor{mycolor}{see the tail of ``S0'' and ``S2'' in Fig. \ref{fig:segms}}). In this case, one must ``jump'' over the \textcolor{mycolor}{critical curve} to connect a close pair of images. The condition is that the close image pair must have the same source position, and their magnifications have a comparable absolute value but with opposite sign. In practice, the ``connecting'' procedure is repeated until all the image segments are closed.



Once the connecting procedure is complete, we obtained closed image boundaries from previously un-closed true solution segments\footnote{An animation shows how closed image boundaries are linked can be visualised at:
\href{https://github.com/rkkuang/triplelens}{https://github.com/rkkuang/triplelens}, together with the source code and documentation.}. And finally, we can move on to calculate the enclosed area using line integral. 

\subsection{Finding \textcolor{mycolor2}{true images} of the lens equation from its complex polynomial}
\label{sec::cri_truesolution}
\textcolor{mycolor}{
For a triple lens system ($\Nlens$=3), the lens equation can be converted into a tenth-order complex polynomial \citep{2002astro.ph..2294R}, which can be solved numerically \citep{skowron2012general}, yielding ten (complex) roots. In most cases, however, not all of these are necessarily \textcolor{mycolor2}{true images} of the original lens equation, i.e., eq. (\ref{lensequ}). In the following, we discuss in more detail how to check whether a root from the complex polynomial roots solver is a true root to the original lens equation.}

\textcolor{mycolor}{
Numerically, the complex roots solver in \citet{skowron2012general} locates roots one by one. At each step, they deflate the original polynomial by the found root, then proceed to search for the next root. The deflation process introduces numerical noise, so they also conduct a ``root polishing'' procedure. This involves taking each root as the initial guess in Newton's (or Laguerre's) method to find a more accurate root of the full polynomial. 
}

\textcolor{mycolor}{
The roots coming from the polynomial solver are not necessarily \textcolor{mycolor2}{true images} of the original lens equation. For a given source position $\zeta$, theoretically a true solution $z$ should satisfy: $\delta = |\zeta - z - f(\overline{z})| = 0$. Nevertheless, due to numerical noise, in practice $\delta$ is not strictly zero even for \textcolor{mycolor2}{true images}. We found that in general, \textcolor{mycolor2}{true images} correspond to $\delta$ $\sim 10^{-16}$ to $10^{-8}$ while \textcolor{mycolor2}{false images} lead to $\delta$ $\sim 10^{-1}$ to $10^{0}$. There is usually a clear separation between \textcolor{mycolor2}{true images} and \textcolor{mycolor2}{false images} in terms of $\delta$. We set the criterion to be $10^{-5}$, i.e., an image is true if it satisfies $\delta < 10^{-5}$.
}

\textcolor{mycolor}{
We note that $10^{-5}$ is not valid in all cases. When the source is just outside cusps or folds, there will be ``nearly true'' false images. Alternatively, if the source is just inside cusps or folds, there will be ``nearly false'' true images. \citet{skowron2012general} implemented their algorithm in double precision. They found for fifth order polynomials, the limiting precision for very close roots is $10^{-7.7}$ for two close roots (near a fold caustic), and $10^{-5.2}$ for three very close roots (near a cusp).
}

\textcolor{mycolor}{
\NewEntry{The situations discussed above do not \textcolor{mycolor3}{affect} our whole method for the following reasons.}{}
\begin{itemize}[topsep=0pt]
    \item Only a small fraction of the source boundary will experience caustic crossing. The probability that discretely sampled points happen to be very close to the caustics is low. Thus only a few points or no points will experience this ambiguity.
    \item \textcolor{mycolor3}{Even if} we miss several true images, we can still link the points into image boundaries using procedures which introduced in \S\ref{sec::conn}. 
    \item If we include some false image points, since they are close to the \textcolor{mycolor2}{true images}, they would not affect the final area significantly.
\end{itemize}
}

\subsection{Criterion for image segment connection}
\label{sec::cri_vicinity}
\textcolor{mycolor}{In \S\ref{sec::conn}, we check whether two segments are connected by checking the distance $d$ between their heads or tails, or the distance between corresponding source positions. We now elaborate further why we choose a threshold of $10^{-5}$. It actually relies on several numerical observations. The first is that although we use $n$ different points to approximate the source boundary, there are actually $n+1$ points for us to use in the code with $\theta_n = \theta_0 + 2\pi$. These two source positions, $\zeta(\theta_0)$ and $\zeta(\theta_n)$, corresponding to exactly the same image points, and they usually correspond to the head and tail of a linked list. Thus the distance $d$ between the head and the tail in this case is exactly zero. The second is when two segments happen to be separated by the critical curves. The segments' head or tail correspond to exactly the same source positions. Finally, in our method, we use this connection check only a few times. It is used after the linked lists are linked and the \textcolor{mycolor2}{false images} are removed. At that time several segments are left. They correspond to either individual boundaries (e.g., ``S7'' in Fig. \ref{fig:segms}), or they are connected at the same points (e.g., ``S0'' and ``S1'' in Fig. \ref{fig:segms}), or they are ``jumping'' over critical curves (e.g., ``S6'' and ``S3'' in Fig. \ref{fig:segms}).
}

\textcolor{mycolor}{
It may seem that the threshold $10^{-5}$ is too large when there are multiple close images, but we note that in a previous step the procedure has already separated different segments from each other. Additionally, in general, image segments which belong to different image boundaries are well separated, or at least their heads and tails are not in the same place. By checking head-tail distance, we will not mix different segments together if they do not belong to the same image boundary.
} \textcolor{mycolor2}{One could set a stricter threshold, e.g., $10^{-10}$ (although for a typical source radius in microlensing events, $10^{-5}$ is already sufficient). 
It is also possible to check for segments connectivity by looking for the nearest loose end (head or tail) of a segment. However, when sampled points are not dense enough, the segments obtained in the previous step may be incorrectly linked (especially when two image boundaries are very close to each other, or when a ring-like structure is formed), and the image areas calculated could be wrong. Taking all these points into account, we choose to use the distance criterion.}

\vspace*{-4mm}
\subsection{Comparison with previous image boundary obtaining methods}
\label{sec::comp}
\textcolor{mycolor}{Previously published papers use different ways to obtain the image boundaries. \textcolor{mycolor2}{\citet{gould1997stokes}, which introduced the contour integration method, included a method for constructing continuous image boundaries.} To avoid inverting the lens equation, \citet{1987contour} proposed the contour plot method to find the image positions. This contour plot method was improved by \citet{dominik1995}. By constructing a squared deviation function and plotting its contour, the image boundaries corresponding to a circular source can be obtained. The overall image information, such as image numbers and shapes, are encoded in the contour plot of the squared deviation function. This method has a limitation: it only provides plots, but not precise numerical values about the image boundaries. To find more precise image contours, one needs high density sampling in the lens plane.}

\textcolor{mycolor}{
\citet{dominik1993, dominik1995} further promoted this idea into a contour-plot-and-correct method. The contour plot data can be stored in purpose designed structures, which can potentially be analysed to obtain microlensing light curves. \citet{dominik1998robust} used the contour plot method to obtain image boundaries and then applied Stokes's theorem to calculate image areas. \textcolor{mycolor2}{In a further development, \citet{2007DominikAdapt} proposed an adaptive contouring algorithm to determine the image contour.}
}


\textcolor{mycolor}{Later works used numerical algorithms to solve the lens equation. In \citet{bozza2010microlensing}, he introduced several error estimators which enable adaptive sampling on the source boundary. He starts with two points on the source boundary, and then inserts new points one by one, between the pair of points which has the largest error estimate. This strategy allows efficient sampling near caustics. Accurate image areas can be obtained with the minimum number of calculations. This method has been developed into the widely used VBBinaryLensing package \citep{bozza2018vbbinarylensing}, which has been integrated into some microlensing events modeling Python packages like pyLIMA \citep{pylima} and MulensModel \citep{mulensmodel}.
}

\textcolor{mycolor}{The caustic structures in triple lenses are much more intricate \citep{2015ApJ...806...99D, 2019ApJ...880...72D} than those for binary lenses, resulting in more complicated image topologies and degeneracies \citep{2014MNRAS.437.4006S}. This poses challenge on obtaining the area of highly distorted images of a source star. Bozza's strategy will require effort to extend it to the triple lens scenario. As a consequence, we have designed and implemented a different approach to determine image boundaries (see \S\ref{sec::conn}).}

\subsection{Stokes' theorem and magnification}
\label{sec::stokes}
\textcolor{mycolor}{
Given any continuous image boundaries, $\left\{ z^{(k)}=x_1^{(k)} + i\,x_2^{(k)}\right\}$, $k=0,1, \cdot\cdot\cdot, n$, where the first and last points are identical,} the enclosed area can be calculated as
\begin{equation}
A =  \frac{1}{2}\sum_{k=1}^{n} (x_2^{(k)}+x_2^{(k-1)}) (x_1^{(k)} - x_1^{(k-1)}),
\end{equation}
or as a more symmetrical expression (\citealt{dominik1998robust}),
\begin{equation}
\begin{split}
A = \frac{1}{4} \bigl[ \sum_{k=1}^{n} (x_2^{(k)}+x_2^{(k-1)}) (x_1^{(k)} - x_1^{(k-1)}) \\
 +(x_1^{(k)} + x_1^{(k-1)}) (x_2^{(k)} - x_2^{(k-1)}) \bigr].
\end{split}
\label{equ:8}
\end{equation}

If there are no nested image boundaries, the magnification is then simply the total area of all the image boundaries divided by the source area. However, there will be nested images in some cases. We handle this by assigning each image boundary object a ``parity'' attribute with value $+1$ or $-1$, according to the sign of magnification at the head of the image boundary. \textcolor{mycolor2}{For example in Fig. \ref{fig:segms}, the heads of ``S1'' and ``S3'' are separated by the critical curve with their magnifications having opposite signs, and thus they are assigned opposite parities.} The total area covered by the source is found by summing up all the boundary areas multiplied by the ``parities''.
\textcolor{mycolor2}{Notice that the (signed) area of each image boundary does not depend on the initial starting point. If we start on ``S1'' with positive parity counter-clockwise, the parity of the boundary will be assigned positive, and the enclosed area calculated with eq. (\ref{equ:8}) will be negative. On the other hand if we start on ``S3'' with negative parity (clockwise), the calculated area will be positive, and so the product of the initial parity with the area calculated with eq. (\ref{equ:8}) will remain the same.}
The total magnification is then simply the sum divided by the area of the source \textcolor{mycolor}{$\pi \rho^2$}. 

The simplest scheme is to start with an approximation of the source limb with e.g., $n=256$ uniformly sampled points, and find the finite source magnification. We can then double $n$, and compare the change in the magnification. We iterate until the relative change is smaller than a preset accuracy, e.g., $\epsilon = 10^{-3}$. However, sampling the source boundary uniformly does not take care of special places on the source boundary, for example, when the source straddles the caustics; these special source boundary places need denser sampling. So we first uniformly sample e.g., $n=45$ \textcolor{mycolor}{different} points on the source boundary, i.e., the $k$-th point \textcolor{mycolor}{$\zeta(\theta_k)$} corresponding to an angle $\theta_k = 2\pi k/n$, \textcolor{mycolor}{with $\theta_0 = 0,\,\theta_n = 2\pi$}. For each \textcolor{mycolor}{$\theta_k$}, we compute the point source magnification \textcolor{mycolor}{$\mups(\zeta(\theta_k))$}, which controls the density of points to be sampled around \textcolor{mycolor}{$\zeta(\theta_k)$}. In this way, we will get an initial sample on the source boundary which takes special care around high magnification places.




\subsection{Bozza's Quadrupole test in deciding whether finite source computation is necessary}
\label{sec::quad}
Since there is no fundamental difference between binary and triple lens system, we adopt the quadrupole test as introduced in \citet{bozza2018vbbinarylensing}, to detect that whether the source star is close to the caustics, and decide if it is necessary to use finite source computation. If it is not necessary, we use point source magnification $\mu_{PS}$ as an approximation.

The finite source magnification of a uniform brightness source can be expanded as \citep{pejcha2008extended}
\begin{equation}
\label{FSexpand}
    \mufs = \mups + \frac{\rho^{2}}{8}\Delta\mups + \frac{\rho^{4}}{192}\Delta^2\mups + O(\rho^5),
\end{equation}
where $\Delta = \frac{\partial^2}{\partial x^2} + \frac{\partial^2}{\partial y^2}$ is the Laplacian and $\Delta^2 = \Delta \Delta$ is the biharmonic operator.
The quadrupole term in equation (\ref{FSexpand}) for each image $z_I$ can be written as (Bozza 2018)
\begin{equation}
\label{muQI}
    \mu_{Q_I} = -\frac{ 2\,{\rm Re}[3\overline{f}^{\,\prime 3}f^{\,\prime \prime 2} - (3-3J+J^2/2)|f^{\prime \prime}|^2+J\overline{f}^{\,\prime 2}f^{\prime \prime \prime}] }{J^5}\rho^2,
\end{equation}
where $f^\prime(z)=df/dz$, $f^{\prime\prime}(z)=d^2 f/dz^2$ and
$f^{\prime\prime\prime}(z)=d^3 f/dz^3$.

To detect the cusp caustic, \citet{bozza2018vbbinarylensing} also constructed an error estimator
\begin{equation}
    \label{errcusp}
    \mu_{C} = \frac{6\,{\rm Im}[\overline{f}^{\,\prime3}f^{\,\prime \prime 2}]}{J^5}\rho^2.
\end{equation}
Thus the condition in the quadrupole test can be written as
\begin{equation}
    \sum\limits_{I}c_Q(|\mu_{Q_I}|+|\mu_{C_I}|) < \delta,
    \label{equ:quadcond}
\end{equation}
where $c_Q$ and $\delta$ are to be chosen empirically so as to make sure there is enough safety margin, in our code, $c_Q=1,\; \delta = 10^{-6}\sim 10^{-2}$, similar to those chosen in \citet{bozza2018vbbinarylensing}.

\subsection{Limb-darkening}
\label{sec::limb}
\textcolor{mycolor}{
In practice, precise modelling of observed light curves needs to include limb-darkening. The linear profile is a reasonable approximation to the limb-darkening for most stars \citep{1921Milne}
}


\begin{equation}
I(r) = \overline{I}f(r),\;\; f(r)=\frac{3}{3-u}\left[ 1-u(1-\sqrt{1-r^2})  \right],
\label{equ:limbd}
\end{equation}
\textcolor{mycolor}{
where $r=\rho_i/\rho$ is the fractional radius at a certain radius $\rho_i$ to the source radius $\rho$, and $\overline{I}$ is the average surface brightness. $u$ is the limb-darkening coefficient. It relates to the $\Gamma$ convention limb-darkening law \citep{2002An}
}
\begin{equation}
I(\vartheta) = \overline{I} \left[ (1-\Gamma) + \frac{3\Gamma}{2}\cos\vartheta \right],
\end{equation}
\textcolor{mycolor}{
by $u = 3\Gamma/(2+\Gamma)$, and $r = \sin\vartheta$, where $\vartheta$ is the emergent angle, $0\leq \Gamma \leq 1$.
}

\textcolor{mycolor}{
We choose the method introduced by \citet{bozza2010microlensing} where they use annuli to approximate the source, summing up the magnification in each annulus weighted by the limb-darkening profile. Error estimators can also be constructed, which allow adaptive sampling of the source profile. If an annulus has the maximum error, it can be split into two sub-annuli. The dividing radius is chosen to equipartition the cumulative function, defined as}
\begin{equation}
F(r) = 2\int_0^rxf(x)dx.
\end{equation}


\textcolor{mycolor}{
We note that \citet{dominik1998robust} introduced a different method to calculate the magnification of a limb-darkened source star, involving two-dimensional numerical integration.
}


\section{Results}

\subsection{Light curves}
\label{sec:lkv}
\begin{figure}
\includegraphics[width=\linewidth]{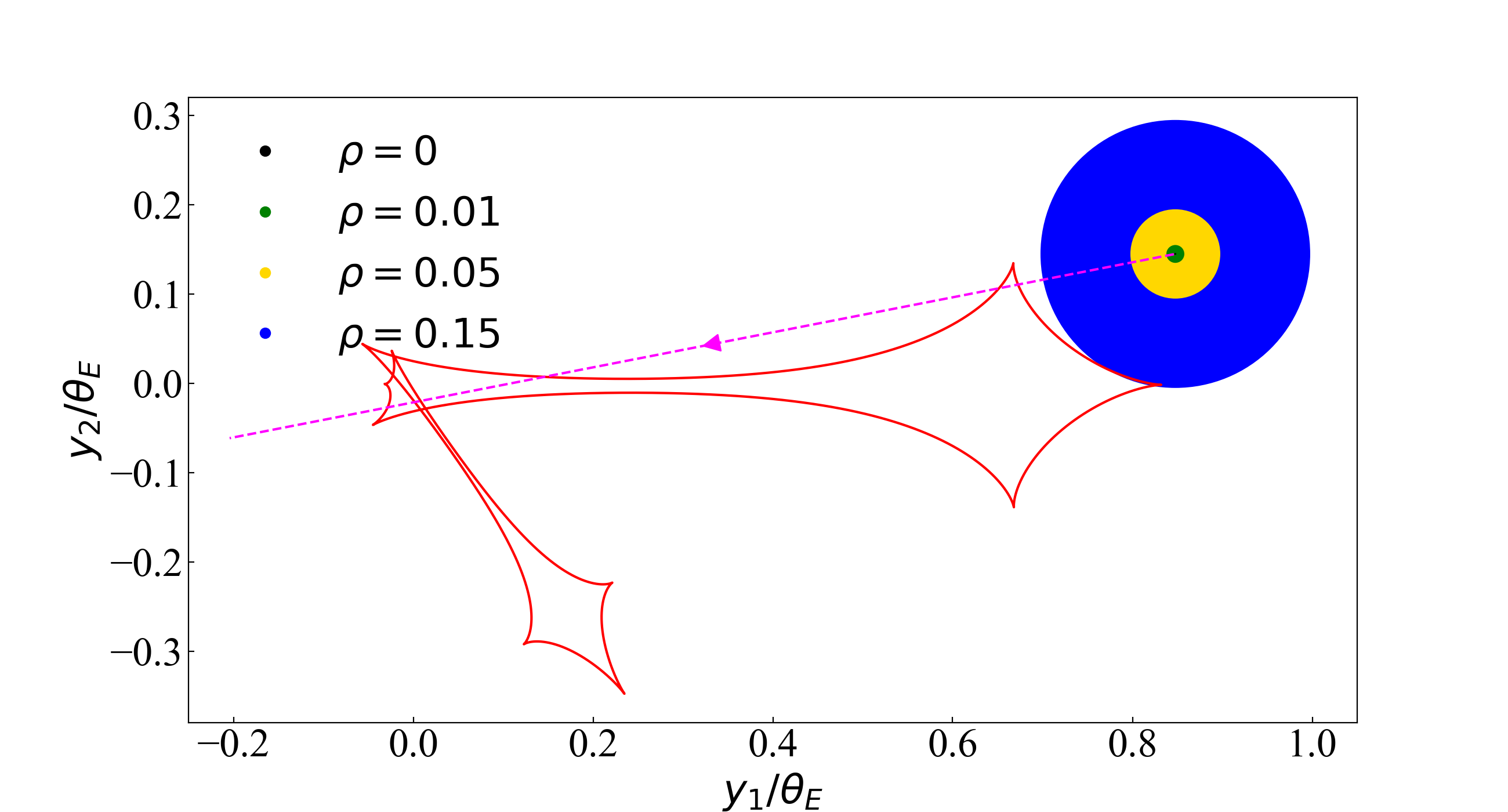}
\vspace{-0.5cm}
\caption{The dashed magenta line shows the source trajectory for which the light curves will be shown in Fig. \ref{fig:lkv} for four source sizes indicated by three circles and a dot (for a point source) at the upper right. The red curve shows the caustics.}
\label{fig:lkv_geo}
\end{figure}

\begin{figure*}
\includegraphics[width=\linewidth]{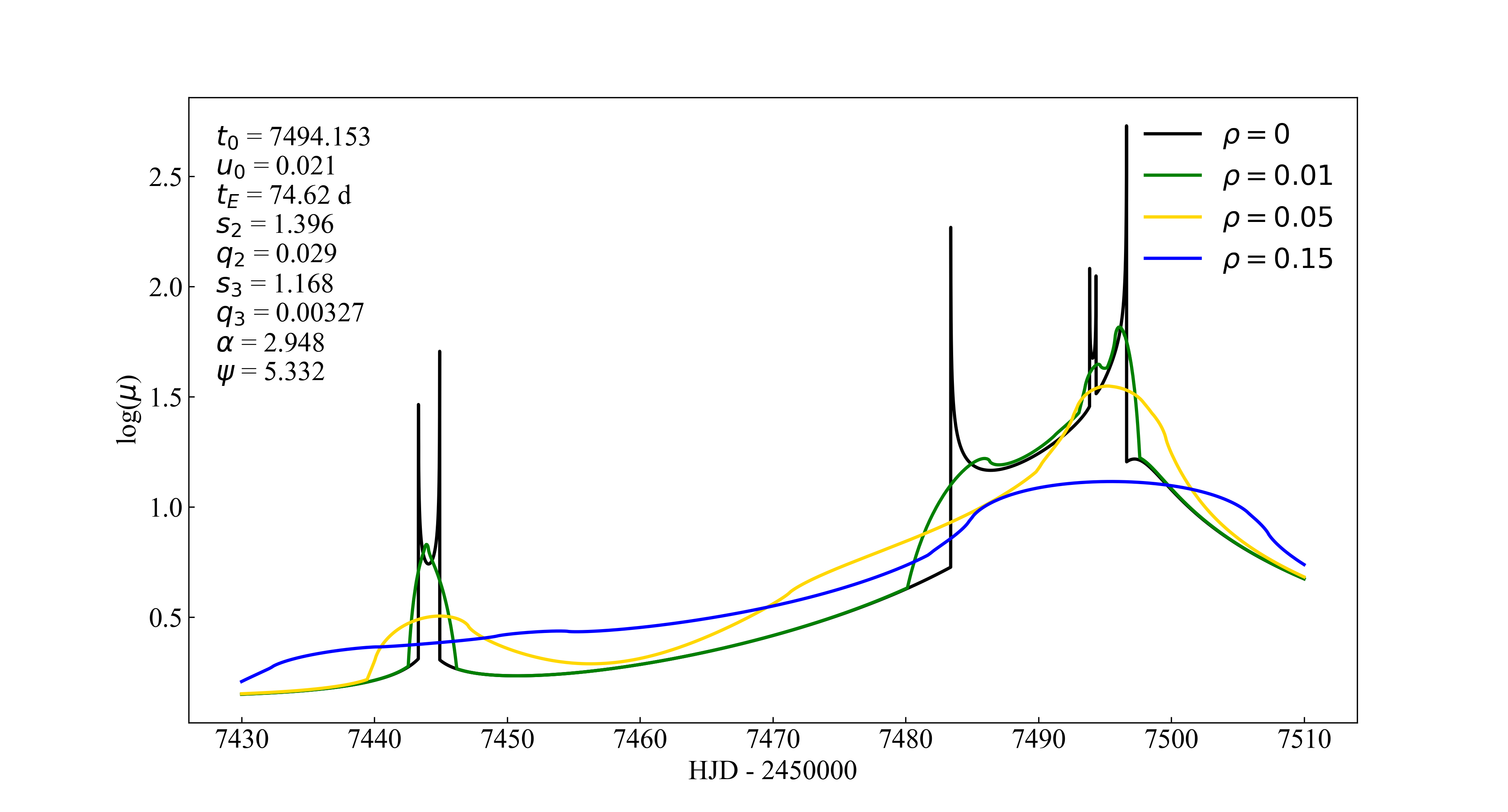}
\vspace{-0.5cm}
\caption{Light curves corresponding to the trajectory indicated in Fig.  \ref{fig:lkv_geo}
for four source sizes, $\rho = 0, 0.01, 0.05, 0.15$ (black, green, gold, and blue). The time starts from HJD $-$ 2450000 $=$ 7430 to HJD $-$ 2450000 $=$ 7510, and $\mu$ is the magnification. Notice that as the source size increases, the values of magnification peaks usually decrease. 
}
\label{fig:lkv}
\end{figure*}

We show several examples of light curves, the triple lens parameters (other than $\rho$) are the same as in \S \ref{sec:topo}. For the trajectory as shown in Fig.  \ref{fig:lkv_geo}, their corresponding light curves are shown in Fig. \ref{fig:lkv} for four source sizes. As the source size increases, the values of magnification peaks are more smoothed out compared to the point source case, and the number of magnification peaks may differ for different source sizes.

\begin{figure*}
\includegraphics[width=\linewidth]{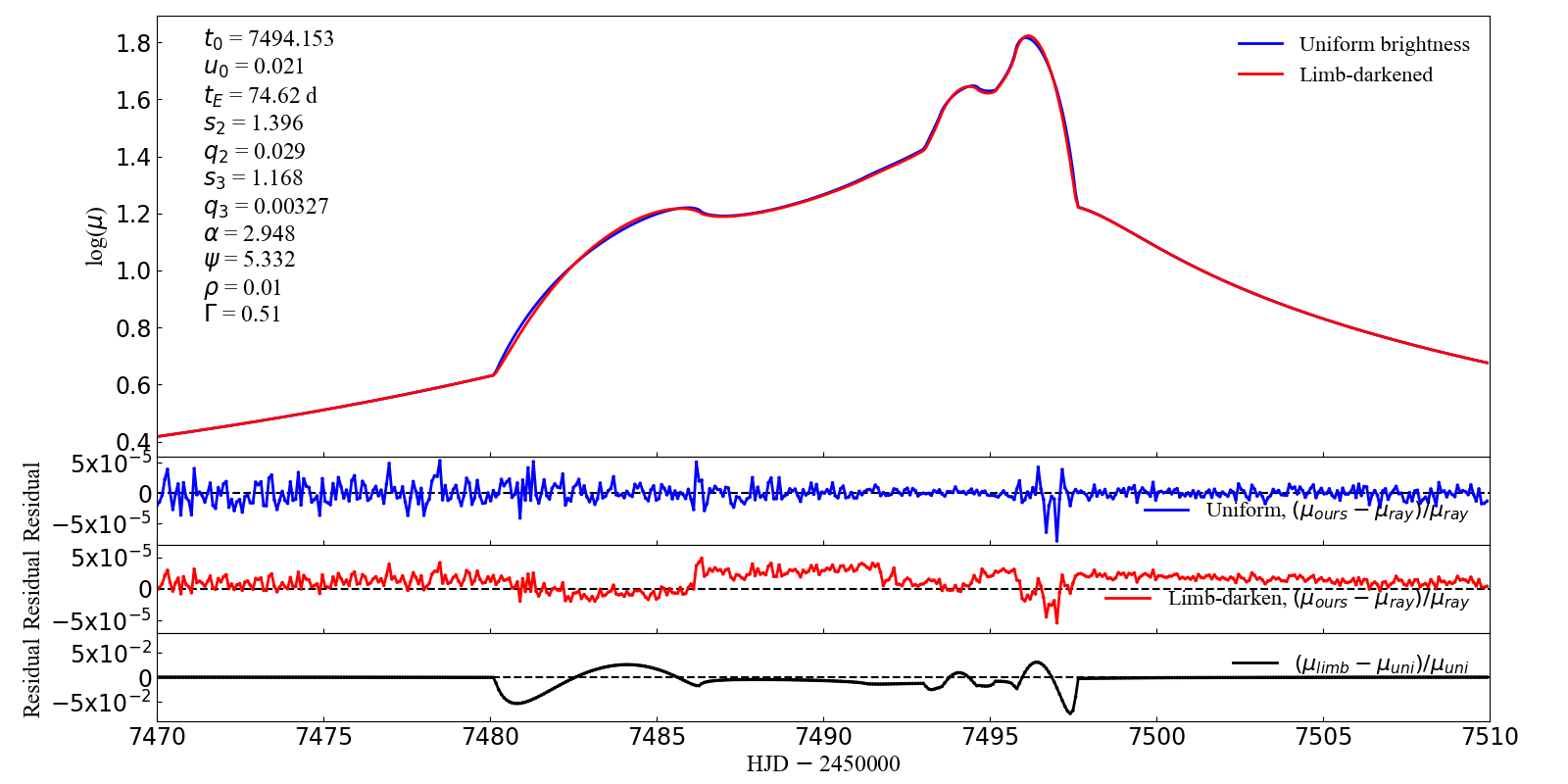}
\vspace{-0.5cm}
\caption{Example light curves for the green source in Fig. \ref{fig:lkv_geo}. $\rho = 0.01$. The time starts from HJD $-$ 2450000 $=$ 7470 to 7510, and $\mu$ is the magnification. The top panel shows \textcolor{mycolor2}{two light curves calculated using our method for the uniform brightness (blue) and limb-darkened (red) cases. The second and third panels show the relative error of magnification of our method ($\mu_{\text{ours}}$) to the result from ray-shooting ($\mu_{\text{ray}}$) for the uniform brightness and limb-darkened cases, respectively.} In both uniform brightness and limb-darkened cases, relative errors of magnification are $\sim 5\times 10^{-5}$. The bottom panel shows the residual of magnification of limb-darkened star to the magnification of uniform brightness star, i.e., $(\mu_{\text{limb}}-\mu_{\text{uni}})/\mu_{\text{uni}}$.
}
\label{fig:lkv_limb}
\end{figure*}

\textcolor{mycolor}{
We show an example of a light curve of a limb-darkened source in Fig. \ref{fig:lkv_limb}. The source radius $\rho = 0.01$, the limb-darkening coefficient $\Gamma = 0.51$. To test our method, we compare our results with those from ray-shooting. The rays are uniformly generated in the lens plane, with $1.6\times 10^9$ rays per $\thetaE^2$. Thus, without lensing, there will be $\sim 5\times 10^5$ rays inside the source boundary. We make comparisons both with and without limb-darkening. The top panel of Fig. \ref{fig:lkv_limb} shows \textcolor{mycolor2}{two light curves calculated using our method for the uniform brightness (blue) and limb-darkened (red) cases. The second and third panels show the relative error of magnification of our method ($\mu_{\text{ours}}$) relative to the result from ray-shooting ($\mu_{\text{ray}}$) for the uniform brightness and limb-darkened cases, respectively.}
In both cases, the relative errors are $\sim5 \times 10^{-5}$. The bottom panel shows the residual of magnification of limb-darkened star to the magnification of uniform brightness star using ray-shooting results, i.e., $(\mu_{\text{limb}}-\mu_{\text{uni}})/\mu_{\text{uni}}$, which shows that the limb-darkening deviation mainly happens during caustics crossing. For example, during caustic entrance (HJD $-$ 2450000 $\sim$ 7480), the limb of the star intersects with the caustic, and the magnification for limb-darkened star is less than the magnification for uniform brightness star. Because from the centre to the edge of a limb-darkened star, the surface brightness decreases gradually.
}

\textcolor{mycolor}{
Each light curve in Fig. \ref{fig:lkv_limb} contains 500 points. If no limb-darkening is involved, it takes $\sim 1$ CPU minute to calculate the light curve using our code. If limb-darkening is considered, it takes $\sim 20$ CPU minutes. This is due to the need to use $15$-$30$ annuli to reduce the modelling error of limb-darkening light curves to be $\sim 5\times 10^{-5}$. In practice, the computing speed can be adjusted by changing the accuracy goal, and by avoiding unnecessary calculations, e.g., during HJD $-$ 2450000 $=$ 7470 to HJD $-$ 2450000 $=$ 7480 in Fig. \ref{fig:lkv_limb}, limb-darkening calculations are unnecessary.
}

\subsection{Magnification maps}
\textcolor{mycolor}{Since the strategy we adopted to model limb-darkening light curves is based on the magnification calculation of uniform brightness stars, we compare the magnification map from our method with both ray-shooting and VBBinaryLensing results for a uniform brightness star.}
\subsubsection{Comparison with ray-shooting}
\begin{figure*}
   \centering
   \includegraphics[width=\linewidth]{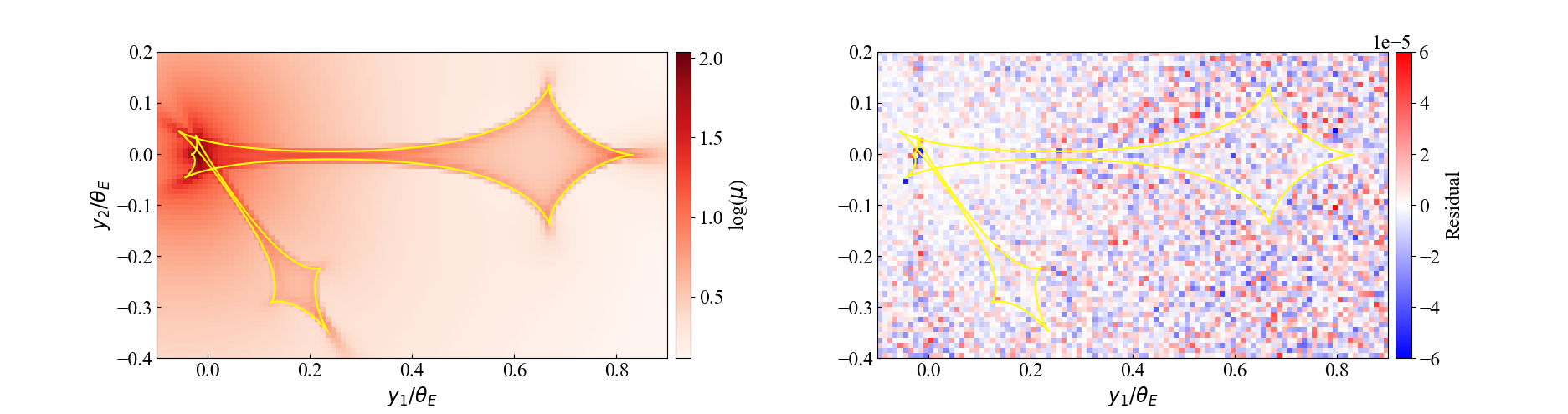}
  \vspace{-0.5cm}
      \caption{Triple lens magnification ($\mu$) map \textcolor{mycolor}{of a uniform brightness star} generated by our method (left panel), and the relative error \textcolor{mycolor}{of magnification} between the results from our method and ray-shooting (right panel), the source radius is \textcolor{mycolor}{$\rho = 0.01$. The \textcolor{mycolor2}{yellow curves in both panels show the caustics}. The pixel size of each map is $100\times 60$. The colourbar scale goes from \textcolor{mycolor2}{$-6\times 10^{-5}$ to $6\times 10^{-5}$} in the right panel, the maximum error (absolute value) is \textcolor{mycolor2}{$5.9\times 10^{-5}$}.} The triple lens parameters (other than $\rho$) are the same as in \S \ref{sec:topo}. }
         \label{fig:map}
   \end{figure*}

We show one magnification map generated using our method, and compare it with one generated using ray-shooting to test the accuracy of our computation. The triple lens parameters (other than $\rho$) are the same as in \S \ref{sec:topo}. Since the ray-shooting method is more suitable for a large source radius, we choose \textcolor{mycolor}{$\rho = 0.01$. The number density of rays shot is \textcolor{mycolor2}{$2.95\times 10^9$} per $\thetaE^2$}. The left panel of Fig. \ref{fig:map} shows the magnification map generated by our method. The right panel of Fig. \ref{fig:map} shows the relative error of magnification compared with the ray-shooting result, which is of the order of $10^{-4}$, \textcolor{mycolor}{with the maximum relative error of magnification (absolute value) being \textcolor{mycolor2}{$5.9\times 10^{-5}$}.}
   
In generating our magnification maps, \textcolor{mycolor}{obtaining polynomial coefficients from lens equations costs 24$\%$ of computation time, 59$\%$ for solving complex polynomials, 4.7$\%$ for initial sampling on the source boundary, and 2.7$\%$ for checking whether a solution is true. The rest part (9.6$\%$) of the computation time is mainly spent on obtaining continuous image boundaries.}
   
\subsubsection{Comparison with VBBinaryLensing}
We have applied our code to the binary lens case, and compared our code to the VBBinaryLensing package \citep{bozza2018vbbinarylensing}. \textcolor{mycolor2}{We used the MulensModel\footnote{\href{https://github.com/rpoleski/MulensModel}{https://github.com/rpoleski/MulensModel}} to calculate the full finite source magnification. To obtain a suitable run for comparison purpose, the accuracy goal (VBBL.Tol) is set to be $ 10^{-5}$.}

\begin{figure*}
\centering
\includegraphics[width=\linewidth]{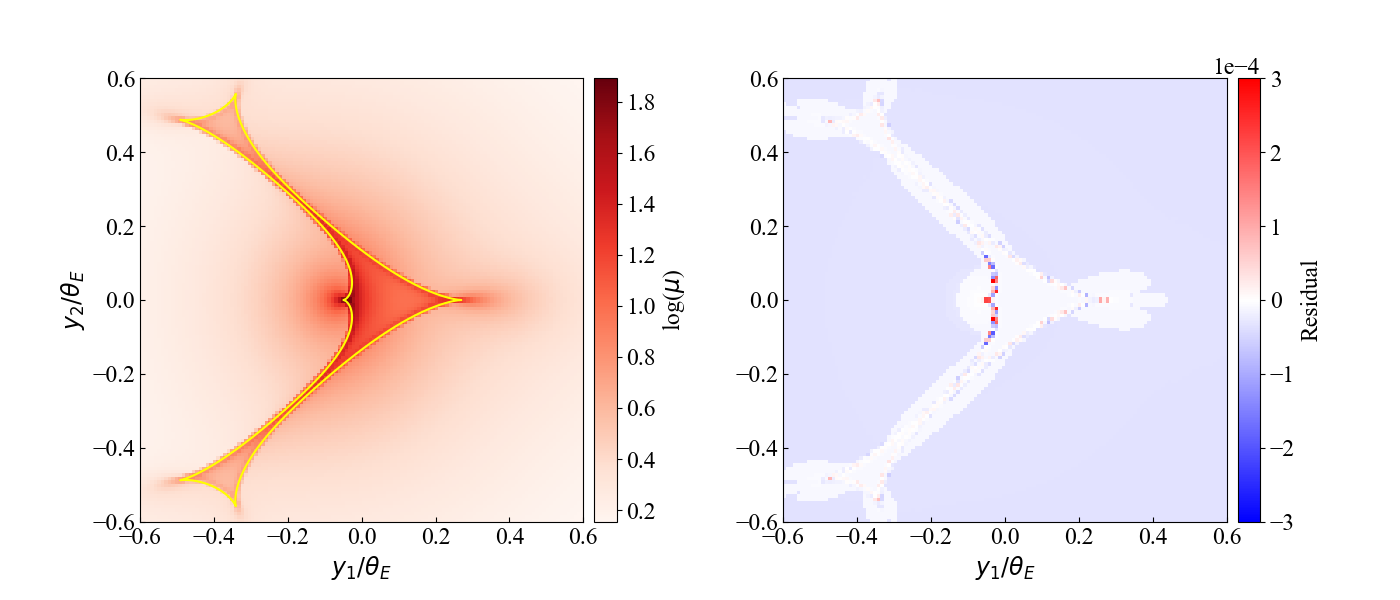}
\vspace{-0.7cm}
  \caption{Binary lens magnification ($\mu$) map \textcolor{mycolor}{of a uniform brightness star} generated by our method (left panel), and the relative error \textcolor{mycolor}{of magnification} between the results from our method and the VBBinaryLensing package (right panel). Binary lens separation $s = 0.8$, mass ratio $q = 0.1$, and source radius \textcolor{mycolor}{$\rho = 0.01$. The yellow curve in the left panel shows the caustics. The pixel size of each map is $128\times 128$. The colorbar scale goes from $-3\times 10^{-4}$ to $3\times 10^{-4}$ in the right panel, the maximum relative error (absolute value) is $3.1\times 10^{-4}$.} \textcolor{mycolor2}{VBBinaryLensing package is used to calculate the finite source magnification as a baseline. Other than the region close to the caustics, our code uses the point source approximation, and the relative error of this approximation is $\sim -3\times 10^{-5}$ .}}
     \label{fig:mapVBBL}
\end{figure*}

We choose the binary lens separation $s = 0.8$, mass ratio $q = 0.1$, and source radius \textcolor{mycolor}{$\rho = 0.01$}. The results are shown in Fig. \ref{fig:mapVBBL}. The left panel shows the magnification map generated by our method, and the right panel shows the relative error of magnification compared with the result from the VBBinaryLensing package. \textcolor{mycolor}{We note that, although both software packages use contour integration to obtain the image areas, this does not imply that the resultant magnification maps will be exactly the same. Although we both use polygons to approximate source boundaries as well as image boundaries, our sampling strategies are different. In addition, the stopping criteria of Bozza’s algorithm is more optimized, being controlled by error estimators. Even though the results are different, the maximum relative error is $3.1\times 10^{-4}$.} Overall, we find our magnification calculating code to be slower by a factor of $\sim 15$ compared to Bozza's package for binary lens systems. This is mostly due to more efficient sampling of points on the limb of source star according to error estimators in their package.

\subsection{High cadence light curves with adaptive sampling}
In our method, which based on Stokes' theorem, the light curve calculation is time-consuming since we have to solve the lens equation many times and need many points to connect the image boundaries. To remedy the situation, we introduce another refinement to speed up the light curve calculation with adaptive sampling.

As shown in Fig. \ref{fig:lkv}, the light curve of a typical event is globally smooth except when approaching/crossing \textcolor{mycolor}{the caustic}. Thus, we can sample the most important points (usually places with large \textcolor{mycolor}{slopes}) in the light curve adaptively, and perform interpolation elsewhere to obtain the full light curve. In this way, we can reduce the number of finite source computations substantially. 

The adaptive sampling procedure is performed as follows: we first compute the magnification $A_1, A_2$ at points $p_1,p_2$, we then compute the magnification $A_c$ in the mid-point, $p_c$, and compare with $A_{\rm mid} = (A_1+A_2)/2$. If the difference between $A_c$ and $A_{\rm mid}$ is larger than a threshold, e.g., $\epsilon = 5\times 10^{-4}$, then we add $p_c$ to our sampled points and repeat the procedure for $p_1, p_c$ and $p_c, p_2$. This process is stopped until the error is smaller than $\epsilon$.

\begin{figure*}
   \centering
   \includegraphics[width=\linewidth]{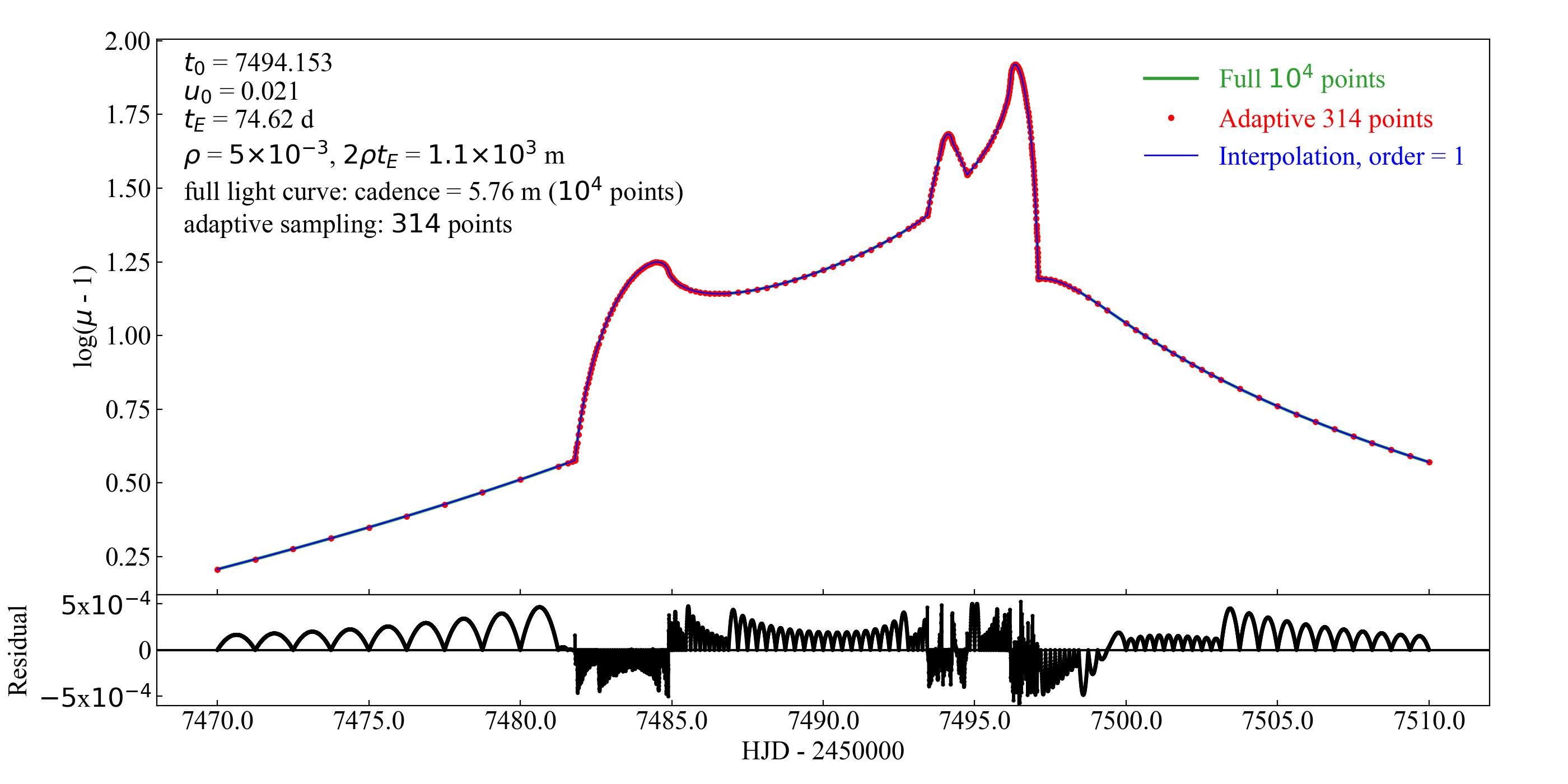}
   \vspace{-0.6cm}
      \caption{Result of adaptive sampling of the light curve (top panel), the full light curve (green solid line) uniformly sample $10^4$ points, corresponding to a sampling of $5.76\times 10^{4}$ minutes. By adaptive sampling, merely 314 points (red dot) are sampled, and the linear interpolated light curve (blue solid line) deviate from the full light curve within a relative error $5\times 10^{-4}$ (bottom panel), $\mu$ is the magnification.}
         \label{fig:adalkv}
   \end{figure*}

Fig. \ref{fig:adalkv} shows one example of this adaptive sampling procedure. The full light curve is uniformly sampled with $10^4$ points, while for the adaptive sampled light curve only 314 points are needed. With linear interpolation, we recover the full light curve with a relative error $\sim 5\times 10^{-4}$ and the CPU time is reduced by a factor of $\sim 3$. Notice that the speedup is not simply the ratio of the number of points ($ 10^4/314 \approx 30$). This is because the adaptively sampled points are mostly at places with large \textcolor{mycolor}{slopes} (e.g., when the source is near caustics) in the light curve, where more time is needed to calculate their magnifications with the finite source effect. \textcolor{mycolor}{We note that, higher order spline interpolations converge faster globally, yet at the entrance and exit of caustics, the light curve is steeper and higher order interpolation yields oscillations. This is known as ``Runge phenomenon'' \citep{runge1901empirical}.}

\section{Summary}

Currently modelling of microlensing light curves of triple lens events often uses methods based on perturbation or ray-shooting. \textcolor{mycolor}{We have developed a method based on establishing continuous image boundaries and contour integration to calculate triple microlensing light curves. Before this work, the contour integration method has not been developed beyond binary lens systems. We first implemented a procedure to obtain the magnification of a source star with uniform brightness, and then extended the procedure to handle limb-darkening.} It is efficient for small source sizes, and complements the ray-shooting method. \textcolor{mycolor}{
Our approach has two advantages: 1) Unlike the contour plot method, we obtain \textcolor{mycolor2}{image boundaries} accurately from solving lens equations. 2) It starts from successively sampled points on the source boundary, which corresponds to successive image tracks. From these image tracks, which contain both true and \textcolor{mycolor2}{false images}, we identify the true image boundaries and calculate the enclosed areas. Our method is a general method which can be applied to any multiple lens system, not just to triple lens systems.
} Ray-shooting is efficient for big finite source sizes, due to Poisson noise in the number of rays each pixel collects. Our independent modelling code is available to be used for cross-checking with other methods.

We have tested our method on light curves and magnification maps, and compared the results with the ray-shooting method. Due to the need to connect continuous closed image boundaries, our method requires more CPU time when the source radius is large (e.g., $\rho = 0.01$). We have implemented an adaptive sampling scheme in the light curve calculation. This may be relevant, for example for the KMTNet \citep{kim2010technical, atwood2012design, kim2016kmtnet}: its cadence can be as short as 10 minutes ($10m/t_{\rm E}\sim 3\times 10^{-4}$ for $t_E=20$ days). In such cases, there is no need to calculate the magnification for every epoch. In practice, the speedup will likely depend on the source size and the number of data points in the light curve.


Our code is publicly available \footnote{\href{https://github.com/rkkuang/triplelens}{https://github.com/rkkuang/triplelens}}, including both the C$++$ source codes and materials to build a Python module ``TripleLensing'' to call the C$++$ codes from Python scripts. We plan to improve further the efficiency of our code and apply it to real triple lens events.

\section{Acknowledgements}
\textcolor{mycolor}{We thank the anonymous referee for providing \textcolor{mycolor2}{many} helpful comments and suggestions.} We thank Valerio Bozza, Jan Skowron, and Andrew Gould for making their codes publicly available. This work is partly supported by the National Science Foundation of China (Grant No. 11821303, 11761131004 and 11761141012 to SM).

\section{Data Availability}
The data generated as part of this project may be shared on a reasonable request to the corresponding author.

\bibliographystyle{mnras}
\bibliography{triple} 






\bsp	
\label{lastpage}
\end{document}